\documentclass[11pt,twoside,epsf]{article}
\usepackage{newpasp}
\def\edcomment#1{\iffalse\marginpar{\raggedright\sl#1\/}\else\relax\fi}
\reversemarginpar

\begin{document}
\title{Size-Luminosity scaling and Inverse Compton Seed Photons in Blazars}
\author{Markos Georganopoulos and  John G. Kirk}
\affil{Max Planck Institut f\"ur Kernphysik, Postfach 10 39 80, D-69029, Heidelberg, Germany}
\author{Apostolos Mastichiadis}
\affil{ Department of Astronomy, University of Athens, GR 15784, Athens, 
Greece}
\begin{abstract}
We present preliminary results of our work on blazar unification.
We assume that all blazars have a broad line region (BLR) and that the size of the BLR scales with the power of the source
in a manner similar to that derived through reverberation mapping
 in radio quiet active galactic nuclei (AGNs).
Using a self-consistent emission 
model that includes particle acceleration we show that according to this scaling, in weak sources  like MKN 421,
the inverse Compton (IC) scattering losses are dominated by  
synchrotron-self Compton scattering (SSC), 
while in powerful sources, like 3C 279,  they are  dominated by external Compton
(EC) scattering of BLR photons. In agreement with other workers, we show that
even in the powerful sources that are dominated by EC scattering, 
the hard X-ray emission is due to SSC. 
Finally, we show that this scaling reproduces well 
the observed sequence of blazar properties with luminosity.

\end{abstract}

\section{Introduction}

Blazars have been shown to exhibit a  sequence of  properties as a 
function of source power (Fossati et al. 1998).
 As the source power  increases, the emission line luminosity and 
  the ratio of Compton to synchrotron luminosity
increase, while the synchrotron 
peak frequency $\nu_{s}$ and the IC peak frequency decrease.  
Recent multiwavelength studies support this scheme (e.g. Kubo et al. 1998),
although the result that the  ratio of the Compton to synchrotron luminosity
increases with source luminosity suffers from limited statistics, 
and should be only considered tentative.

The initial division of blazars into flat spectrum radio quasars (FSRQs)
and BL Lacertae objects (BLs) was based on the equivalent width (EW) of the
broad emission lines. Sources with EW $>$  5 \AA $ $ were classified as FSRQs
and sources with EW $<$  5 \AA $ $ as BLs. 
This difference in the EW of the emission lines has been interpreted
as absence of a substantial BLR in BLs, and 
has been used to advance the idea that in BLs  the GeV-TeV
emission is due to  SSC, while in FSRQs the GeV emission is
due to EC. 
However, the EW criterion does not correspond to a dichotomy
 (Scarpa \& Falomo 1997), 
because a weak BLR is present
in BLs as well. Additionally, it is not the BLR luminosity $L_B$ 
that is relevant
for the EC luminosity, but the BLR photon energy density $U_{E}$ 
that is measured in 
the comoving frame of the non-thermal source. One needs to know 
not only  $L_B$, but also the BLR radius $R_B$ 
and the location of 
the non-thermal emitter
relative to the BLR in order to quantify the relative importance of SSC versus
EC emission. 

The size of the BLR has  been measured only for radio quiet AGNs through 
reverberation mapping (Kaspi et al. 2000), and it has been found to scale with
the ionizing luminosity $L_{acc}$ of the acrretion disk as $R_B \propto L_{acc}^{0.7}$.
It was additionally found that, assuming Keplerian motions for the BLR clouds,
the mass of the central object scales with the ionizing luminosity as 
$M\propto L_{acc}^{0.5}$.  

Here we assume the scalings apply also to radio loud AGN and examine 
the implications for blazars.
Following the formalism of internal shocks propagating in a conical jet
(Rees 1978),
we scale  the location $D_{blob}$  and the radius $R_{blob}$ of 
the non-thermal spherical emitter  with the mass $M$ of the central
object. We assume that  $L_B$ is a fraction of $L_{acc}$
and scale the kinetic luminosity and the Poynting flux of the blob
  with  $L_{acc}$.

\section{The scaling  and its application}

Under this scheme, the radius of the BLR scales as $R_B\propto L_{acc}^{0.7}$,
and the distance of the blob from the center of the system scales as
$D_{blob}\propto M \propto L_{acc}^{0.5}$. 
Therefore, we assume that for powerful sources the blob  radiates 
from inside the BLR, $D_{blob}<R_B$. However, if we consider increasingly 
weaker sources,  $R_B$ falls  faster than $D_{blob}$ and,  below some
critical luminosity,  the blob  radiates from outside the BLR.
For a blob inside the BLR, $U_{E}\propto (L_B/R_B^2) \Gamma ^2$,
where $\Gamma$ is the bulk motion Lorents factor of the blob.
For a blob outside the BLR, 
the solid angle subtended by the BLR,  as  seen 
by the blob,  is  reduced. 
This affects $U_{E}$ in two ways.
The first is the usual geometric  $1/r^2$ factor, and the second is 
a relativistic de-boosting that, for $D_{blob} \gg R_B$, 
 gives  $U_{E}\propto (L_B/D_{blob}^2) /\Gamma ^2$
(Dermer \& Schlickeiser 1994). The second effect dominates  if  the blob
is located less than 
 a few $R_B$ away from the BLR. For example, for $\Gamma=10$,
$U_E$ at a distance of $10 \; R_B$ is smaller than $U_E$ inside the BLR by
a factor of $10^{2}$ due to the usual $1/r^2$ and by a factor of 
$\approx 10^{4}$ due to relativistic de-boosting.

We apply this scaling  using a  model (Georganopoulos \& Kirk 2000)
for the blob which includes both
particle acceleration and radiative losses with IC scattering losses 
treated in the Thomson regime.
Particles are accelerated in an acceleration zone and escape into a 
radiation zone, before they eventually
escape out of the system.

This simplified picture is intended to 
represent particles which are accelerated
by a shock front, escape it, and radiate downstream before the compressed 
plasma re-expands. Whilst undergoing acceleration, particles simultaneously 
suffer synchrotron losses and losses by Compton scattering of both photons of
external origin (EC) and synchrotron photons (SSC) from the radiation zone. 
In the radiation zone, 
EC, SSC and synchrotron losses occur until the particle finally escapes
after a time $t_{esc}=3 R_{blob}/c$. The electron energy distribution (EED) is computed
self-consistently, taking account of these processes.

Our assumption about the scaling of the Poynting flux implies a constant
magnetic field, which we take  to be  $B=0.2$ G. The blob is  assumed to move
with a Lorentz factor $\Gamma=15$ at an angle $\theta=1/\Gamma$
with respect to the  line of sight. Low energy particles ($\gamma_0=10$) 
are injected into the blob at a rate of $Q=0.1$ cm$^{-3}$ s$^{-1}$.
Also $L_B  =0.01 L_{acc}$.
For $L_{acc}=10^{44}$ erg s$^{-1}$, 
reverberation mapping gives 
 $R_B\approx 8.6 \; 10^{16}$ cm. We assume that at this luminosity $R_B=D_{blob}$ and
$R_{blob}=2 \;10^{15}$ cm.

In Fig.1,  we let $L_{acc}$ vary from $10^{42}$  erg s $^{-1}$  to
 $10^{46}$  erg s$^{-1}$,
 and follow the behavior of the system. In the lower panel 
 $R_B$, $D_{blob}$, and $R_{blob}$ are plotted 
as a function of $L_{acc}$. For weak
sources, the blob is located outside the BLR ($D_{blob} > R_B$). As the source
power increases, the blob gradually approaches  
and eventually enters the BLR. In the middle  panel of Fig.1 we see how
 this affects $U_E$ (the magnetic field energy density $U_B$ is constant under
the adopted scaling). For weak sources, $U_E$ is much smaller than both 
$U_B$ and the self-consistently derived synchrotron photon density $U_S$.
Gradually $U_E$ increases, and eventually, for bright sources, 
dominates over both $U_B$ and $U_S$. 
In the upper panel of Fig.1 we plot the self-consistently derived break 
$\gamma_{break}$ in the EED. Note that  $\gamma_{break}$ decreases as the
power of the source decreases.
\begin{figure}
\plotfiddle{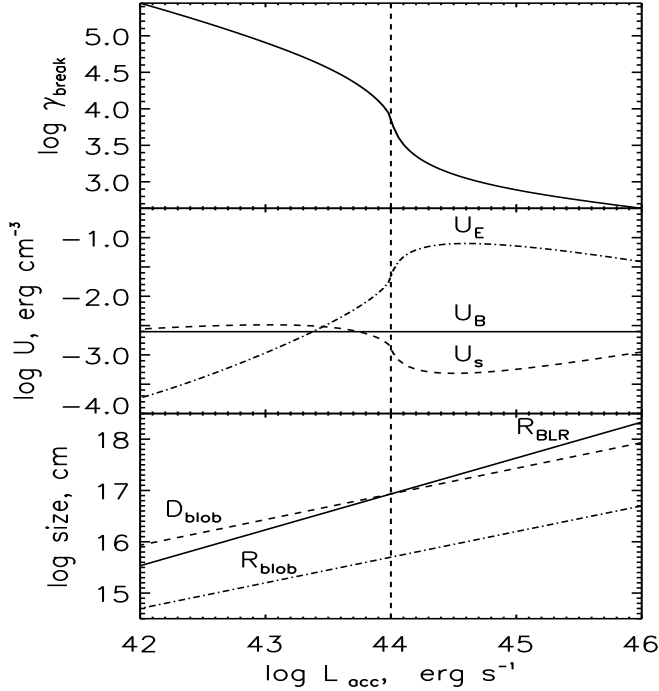}{3.8in}{0.}{45}{35}{-140.}{0.0}
\caption{Sizes (lower panel), energy densities (middle panel), and the break
of the EED as a function of the accretion disk luminosity.}
\end{figure}

In Fig.2 we plot  basic observable quantities as a function of $L_{acc}$.
In the lower panel we plot the synchrotron luminosity $L_S$, the SSC luminosity
$L_{SSC}$, and the EC luminosity $E_{EC}$. At low powers,  $L_{EC}$ is much 
weaker than $L_{SSC}$ and $L_S$ that are roughly equal. As the source power
increases,  $L_{EC}$ gradually dominates over $L_{SSC}$ and $L_S$, and we
end up with an EC dominated source. In the middle panel we plot the peak
frequencies of the three emission components. Note how the synchrotron peak 
frequency $\nu_S$ decreases 
as the source power increases. 
\begin{figure}
\plotfiddle{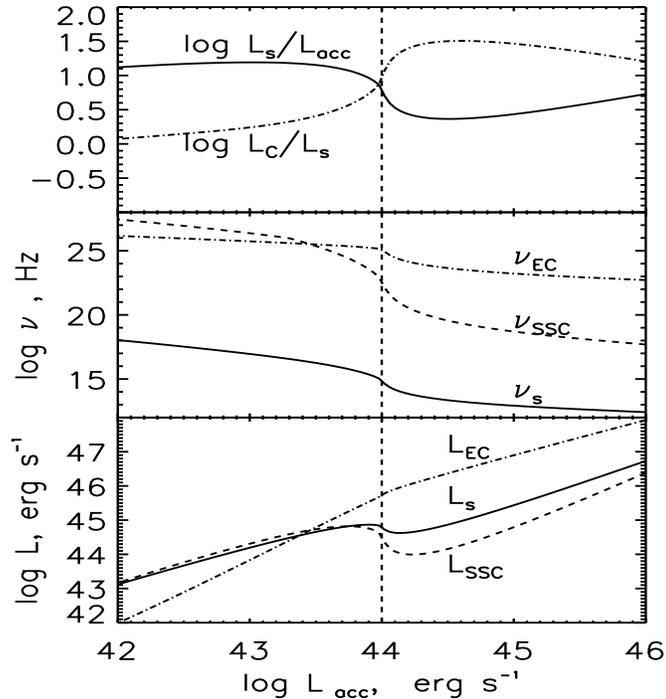}{3.8in}{0.}{45}{35}{-120.}{0.0}
\caption{Luminosities (lower panel), peak frequencies (middle panel) and the 
ratio or synchrotron to accretion disk luminosity and IC to synchrotron luminosity (upper panel).}
\end{figure}
Similar behavior is also seen for the SSC  
peak 
frequency $\nu_{SSC}$ and the EC peak frequency $\nu_{EC}$. 
Finally in the upper panel of Fig.1 we plot the ratio 
$L_S/L_{acc}$ (solid line) of the synchrotron to accretion disk luminosity 
and the ratio $L_{C}/L_{S}$ of the IC luminosity to synchrotron luminosity. 
$L_S$ dominates over $L_{acc}$ in  weak sources in agreement with the lack
of a thermal component and weak emission lines in weak sources like BLs.
$L_{acc}$ becomes more significant for powerful sources, again in agreement
with the strong emission lines of FSRQs and the accretion disk signature
observed in some FSRQs (e.g. in 3C 279, Pian et al 1999).
The dominance of the Compton over the synchrotron emission increases as 
the source power increases, in agreement with  observations 
(e.g. Kubo et al. 1998).
Note, however,  that the observational case for  an increasing Compton dominance with source power 
is based on poor statistics due to the limited sensitivity of the {\it CGRO}.

In Fig.3 we plot the synchrotron luminosity  $L_s$ versus the 
synchrotron  peak frequency   $\nu_s$ for the  blazars studied by  
Sambruna, Maraschi \& Urry (1996) and Kubo et al. (1998). 
On top of the data points we plot the model tracks as a function
of $L_{acc}$   for a range of observing angles. Note that these tracks 
represent two model derived quantities, $L_S$ and $\nu_S$ 
as  $L_{acc}$ varies. The model covers rather well the observed parameter
space. Given the track of the model under an angle $\theta=0^{\circ}$, we
do not expect that any powerful ($L_S\approx 10^{47}$ erg s$^{-1}$) sources
with high peak frequencies $\nu_S\approx 10^{17}$ Hz exist. The discovery of
such sources would pose a serious problem for this model.   
   
\begin{figure}
\plotfiddle{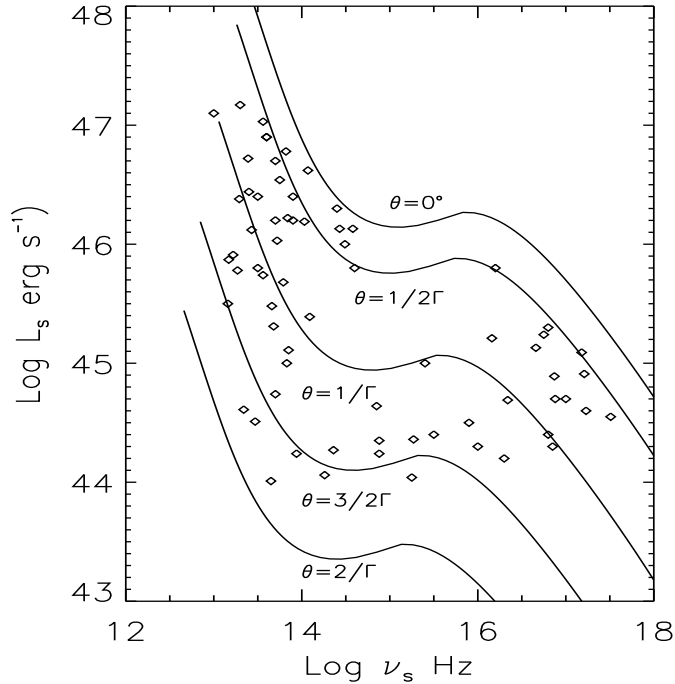}{4.in}{0.}{45}{35}{-120.}{0.0}
\caption{Synchrotron  luminosity versus synchrotron peak frequency for a set
of blazars (Sambruna et al. 1996, Kubo et al. 1998). 
Overlayed are the  the model tracks as a function
of $L_{acc}$   for a range of observing angles.}
\end{figure}

\section{The luminosity scaling }

In Fig.4 we plot the peak luminosities of the three emission mechanisms for
 5 model sources with $L_{acc}$ ranging from $10^{42}$ up to $10^{46}$ erg
s$^{-1}$. The behavior of the model is in good agreement with the general
characteristics of the observed luminosity sequence 
(compare with Fig.12 of Fossati et al. 1998). 
The synchrotron peak frequency decreases from $10^{18}$ Hz down to 
$10^{12}$ Hz, as the  synchrotron power increases. The Compton peak frequency
in weak sources is  in the TeV regime ($\approx 10^{25-26}$ Hz) 
and it is due to SSC, while in  powerful sources it is in the GeV regime 
($ \approx 10^{23}$  Hz) and it is due to EC. According to the model,
although in these bright sources the energy losses are dominated by EC,
  the hard X-ray emission is due to SSC, in agreement 
with resent observations (Kubo et al. 1998).

We note here that Fig.4 corresponds to sources that are oriented at an angle
$\theta=1/\Gamma$. In reality one should expect a range of angles,
which will  give rise to significant  scattering around the presented trend.
This  scattering is visible  in the  work  of 
Fossati et al. (1998) 
 and indicates that a proper unification scheme for blazars
should  include the effects of orientation.

\begin{figure}
\plotfiddle{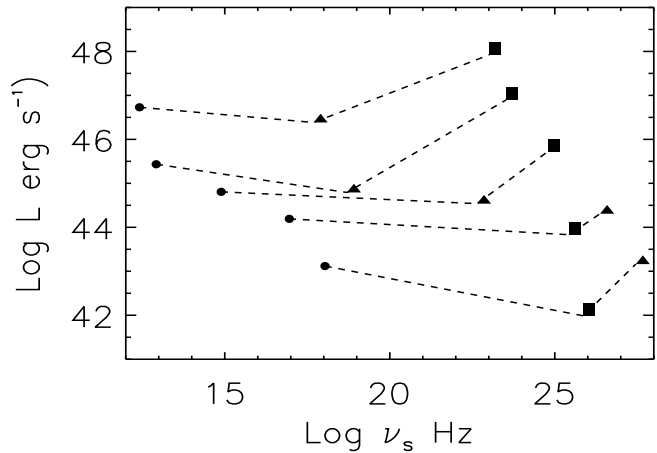}{2.5in}{0.}{45}{40}{-120.}{0.0}
\caption{The model luminosity sequence. The synchrotron (circles), SSC (triangles), and EC (squares) luminosity as a function of frequency for 5 model sources
of different intrinsic power. The broken lines link points from a single model.}
\end{figure}

\end{document}